\documentstyle[psfig,11pt,fullpage]{article}
\begin{document}


\begin{center}
\thispagestyle{empty}
{\Large{\bf Comment on the Calculation of the Angular Momentum
and Mass  for the (Anti-) Self Dual Charged Spinning $BTZ$ Black Hole}}\\
\vspace{1.5cm}
Kevin C.K. Chan\footnote
{\normalsize E-mail address: kckchan@avatar.uwaterloo.ca}\\
Department of Physics, University of Waterloo,\\
Waterloo, Ontario, Canada, N2L 3G1.\\
\end{center}
\vspace{0.5cm}
\centerline{ABSTRACT}
\bigskip

A recent paper [M. Kamata and T. Koikawa, Phys. Lett. {\bf B353} (1995)
196.] claimed to obtain the charged version of the $(2+1)$-dimensional spinning
$BTZ$ black hole solution by assuming a (anti-) self dual condition imposed
on the electric and magnetic fields. We point out that
the angular momentum and mass diverge at spatial infinity
and as a consequence the solution is unphysical

\vspace{0.2 cm}


\noindent{PACS number(s): 04.20.Jb 97.60.Lf}

\vspace{0.5cm}

Recently, Ba\~{n}ados, Teitelboim and Zanelli ($BTZ$) \cite{BTZ,BHTZ} found
that $(2+1)$-dimensional general relativity with a negative cosmological constant
admits a black hole solution which has finite mass $M$ and
angular momentum $J$ identified at spatial infinity.
The authors also mentioned in \cite{BTZ} that an electrically
charged spinning $BTZ$ solution could be obtained.
However, this charged spinning
$BTZ$ black hole is not a solution since it should
have a nonvanishing magnetic field as well.

More recently, Kamata and Koikawa \cite{KK} claimed to extract a charged
version
of the spinning $BTZ$ solution by assuming a (anti-) self dual equation
$E_{\hat r} = \varepsilon B_{\hat{~}}$, $\varepsilon = \pm 1$, where
$ E_{\hat r} \equiv F_{\hat{t}\hat{r}} $ and $ B_{\hat{~}} \equiv
F_{\hat{r}\hat{\phi}}$ are
the orthonormal basis components of the electric field and the magnetic field,
respectively. The Einstein-Maxwell action considered in their paper is
\begin{equation}
S=\int d^3x\sqrt{-g}(R+2\Lambda-F^2),
\label{eq:EMaction}
\end{equation}
where, for simplicity, we have set $4\pi G=1$ and $\Lambda>0$ corresponds
to the anti-de Sitter case.
Putting a stationary and axisymmetric metric in the following form:
\begin{equation}
ds^2=-N^2dt^2+L^{-2}dr^2+K^2(N^{\phi}dt+d\phi)^2, \label{eq:lineelement}
\end{equation}
where $ N $, $ L $, $ K $ and $ N^\phi $ are functions of only r,
their solution to the field equations which follow from (\ref{eq:EMaction})
reads
\begin{eqnarray}
ds^2 &=& -\frac{r^2}{K^2}
\left(-2Q^2 + \Lambda r^2 +\frac{Q^4}{\Lambda r^2} \right) dt^2 +
\frac{dr^2}{-2Q^2 + \Lambda r^2 +\frac{Q^4}{\Lambda r^2}} \nonumber \\ & &
\hspace{1cm} +K^2
\left[ -\frac{\varepsilon Q^2}{\sqrt{\Lambda} K^2} \left(
1+\frac{K^2-r^2}{r_0^2} \right) dt+d\phi \right]
^2,
\label{eq:ds^2 of sol}
\end{eqnarray}
where the radial function is given by
\begin{equation}
K^2= r^2 + r_0^2 \ln|\frac{r^2- r_0^2}{r_0^2}|. \label{eq:K(r)}
\end{equation}
$Q$ is the magnitude of the electric charge and $r_0^2=Q^2/\Lambda$.
Note that $ K^2 $ approaches $ r^2 $ as $ r \rightarrow \infty $.
The authors in \cite{KK} claimed that this line element asymptotically
approaches that of the form of the spinning BTZ solution
\cite{BTZ,BHTZ}.
However, a careful inspection of the line element
indicates that this is not the case, since
the term $K^2-r^2$ in the angular shift function $N^{\phi}$
approaches $r_0^2\ln|\frac{r^2}{r_0^2}|$ instead of zero as
$r\rightarrow\infty$. More precisely, the angular shift function
is identified as
\begin{equation}
N^{\phi}= -\varepsilon \frac{Q^2}{\sqrt{\Lambda}}
\frac{ 1+\ln|\frac{r^2-r_0^2}{r_0^2}|}{K^2}. \label{eq:N^phi}
\end{equation}
The authors compared the asymptotic form of
(\ref{eq:N^phi}) with the formula $N^{\phi}= -J/(2r^2) $
of the BTZ solution \cite{BTZ,BHTZ} and concluded that
their solution has the angular momentum
\begin{equation}
J = \varepsilon \frac{2Q^2}{\sqrt{\Lambda}} \label{eq:AM}
\end{equation}
at spatial infinity.
This is obviously incorrect since as $r\rightarrow\infty$,
the angular shift function in (\ref{eq:N^phi}) approaches
$-\varepsilon\frac{Q^2}{\sqrt{\Lambda}}
\left(\ln|\frac{r^2}{r_0^2}|\right)/r^2$.
It is because the logarithmic term dominates as $r\rightarrow\infty$.
As a result, the angular momentum at spatial infinity should read
\begin{equation}
J = \varepsilon \frac{2Q^2}{\sqrt{\Lambda}}\ln|\frac{r^2}{r_0^2}|
\label{eq:AMa}
\end{equation}
instead of the one given by (\ref{eq:AM}). This fact is further supported
by using the quasilocal formalism
\cite{BY, BCM} to identify the angular momentum in a given solution.
For the metric of the form (\ref{eq:lineelement}), the angular momentum
$j(r)$ at a radial boundary $r$ reads
\begin{equation}
j = \frac{LN^{\phi'}K^3}{N}, \label{eq: quasi}
\end{equation}
where the prime denotes an ordinary derivative with respect to $r$.
$J$ is defined as $J=j(\infty)$. For their solution (\ref{eq:ds^2 of sol}),
we get
\begin{equation}
j = \varepsilon \frac{2Q^2}{\sqrt{\Lambda}}\ln|\frac{r^2-r_0^2}{r_0^2}|.
\label{eq: quasilocal}
\end{equation}
It is trivial to see that as $r\rightarrow\infty$, $j(\infty)$ reduces to
(\ref{eq:AMa}).  In addition, it is worthwhile to calculate the quasilocal mass
by using the quasilocal mass formula developed in \cite{BY, BCM}.
Using the metric (\ref{eq:lineelement}),  the quasilocal mass $m(r)$
at $r$ can be written as 
\begin{equation}
m =
2N\left[L_o\left(\frac{dK}{dr}\right)_o-L\left(\frac{dK}{dr}\right)\right]
- jN^{\phi}. \label{eq: mass}
\end{equation}
The expression inside the square bracket is the quasilocal energy, $E(r)$.
Here $L_o\left(\frac{dK}{dr}\right)_o=g^{KK}_o$ is a background metric component which
determines the zero of the energy. The background can be obtained simply by
setting constants of integration of a particular solution to some special value
that then specifies the reference spacetime. We set $Q=0$ in (\ref{eq:ds^2 of sol}) as the background and as a consequence it is the vacuum anti-de Sitter spacetime. The same background was used in
\cite{BCM} for the calculation of the quasilocal mass of the spinning $BTZ$ black hole.
Now $L_o\left(\frac{dK}{dr}\right)_o = \sqrt{\Lambda}K$. The quasilocal mass
at spatial infinity is defined as  $m(\infty)=M$. As $r\rightarrow\infty$,
$N(r)\rightarrow \sqrt{\Lambda}r$,  $L(r)\rightarrow\sqrt{\Lambda}r\left(1-\frac{Q^2}{\Lambda r^2}\right)$, $\frac{dK}{dr}\rightarrow \left(1-\frac{r_0^2}{2r^2}ln|\frac{r^2}{r_0^2}|\right)$ and $jN^{\phi}\rightarrow 0$. Now it is easy to see that the quas
ilocal energy $E(\infty)$ vanishes but
\begin{equation}
M \rightarrow  2\Lambda\left(\frac{Q^2}{\Lambda} +
r_0^2ln|\frac{r^2}{r_0^2}|\right).
\label{eq:massa}
\end{equation}
Thus $M$ diverges logarithmically, similar to the situation in the angular
momentum. The authors in \cite{KK} loosely compared their solution
(\ref{eq:ds^2 of sol}) and (\ref{eq:K(r)}) with the $BTZ$ case and concluded that the mass is finite and given by $M=2Q^2$. Obviously, they missed the logarithmic term on the right hand side of (\ref{eq:massa}).

Although the angular shift function in (\ref{eq:N^phi}) vanishes
as $r\rightarrow\infty$, it is not a sufficient condition for a
finite $J$. One can deduce from (\ref{eq: quasi}) that for
a given spinning solution $N^{\phi}$ must vanish fast enough at spatial
infinity in order for the solution to admit a finite $J$.
As discussed by the authors in \cite{KK}, the origin of the ``angular
momentum'' in their solution (\ref{eq:ds^2 of sol}) can be considered to be the ``Poynting
pseudovector'' $ E_{\hat r}B_{\hat{~}} $, which gives the black hole the
nonzero angular momentum. One may suspect that it is
the (anti-) self dual condition which leads to a diverging $J$, and
if one relaxes this condition, it should be possible to obtain a general
charged spinning solution with a finite $J$.
However, it was mentioned in \cite{hw} that the quasilocal mass of the static electrically charged $BTZ$ black hole also diverges logarithmically. This ``intrinsic'' logarithmic divergence in mass and angular momentum is due to the fact that the ``electro
static potential'' term in $g_{tt}$ and $g_{rr}$ is logarithmic (a $2+1$ electric point charge actually represents a linear charge
density in $3+1$ dimensions). Even if one relaxes the (anti-) self dual condition, the divergence in angular momentum and mass may still persist in any solution to the field equations of (\ref{eq:EMaction}) with non-vanishing Maxwell's fields. 

Finally, it is interesting to note that a static magnetic solution to the field equations of (\ref{eq:EMaction}) is extracted in \cite{hw}. The solution reads
\begin{eqnarray}
ds^2 &=& -\Lambda(\rho^2+{\tilde r}^2-r_+^2)dt^2+\left(\rho^2+Q_m^2ln|1+\frac{\rho^2}{{\tilde r}^2-r_+^2}|\right)d\phi^2
\nonumber \\ & &
\hspace{1cm}+\frac{\rho^2d\rho^2}{\Lambda\left(\rho^2+{\tilde r}^2-r_+^2\right)(\rho^2+Q_m^2ln|1+\frac{\rho^2}{{\tilde r}^2-r_+^2}|)},
\label{eq:magnetic}
\end{eqnarray} 
where $Q_m$ is the magnitude of the magnetic charge, $r_+$ and ${\tilde r}$ are integration 
constants satisfying ${\tilde r}^2+Q_m^2ln|\Lambda({\tilde r}^2-r_+^2)|=0$. $\rho$ is the usual radial co-ordinate. When $Q_m=0$, (\ref{eq:magnetic}) exactly reduces to the
uncharged $BTZ$ solution. This magnetic solution is free from event horizons and curvature singularities. 
There is a logarithmic term in  $g_{\rho\rho}$ and $g_{\phi\phi}$. One may expect that the quasilocal mass should be divergent as
in the electrically charged $BTZ$ case. Although the quasilocal mass is not explicitly calculated in \cite{hw}, it can be evaluated at spatial infinity by using (\ref{eq: mass})\footnote{The author thanks E. W. Hirschmann and D.L. Welch for drawing his at
tention to a correct calculation.}. Again, we use the vacuum anti-de Sitter spacetime as the background by setting $r_+=0$ and $Q_m=0$. Now $L_o\left(\frac{dK}{d\rho}\right)_o=\sqrt{\Lambda}K$.  As $\rho\rightarrow\infty$, $N(\rho)\rightarrow\sqrt{\Lambda
}\rho$, 
$L(\rho)\rightarrow\sqrt{\Lambda}\rho\left(1+\frac{{\tilde r}^2-r_+^2}{2\rho^2}\right)\left(1+\frac{Q_m^2}{2\rho^2}ln|\frac{\rho^2}{{\tilde r}^2-r_+^2}|\right)$, and  $\frac{dK}{d\rho}\rightarrow\left(1-\frac{Q_m^2}{2\rho^2}ln|\frac{\rho^2}{{\tilde r}^2-r
_+^2}|\right)$. It can be checked that although the quasilocal energy vanishes at spatial infinity, the mass is given by
\begin{equation}
M \rightarrow \Lambda \left( Q_m^2 ln|\frac{\rho^2}{{\tilde r}^2-r_+^2}|+r_+^2-{\tilde r}^2\right)
\label {eq: magmass} 
\end{equation}
which is diverging for $Q_m\neq 0$ (if $Q_m=0$, $M=\Lambda r_+^2$). 
Thus the logarithmic divergence still persists in the pure magnetic case. 
The problem of the divergence may be cured by adding a
topological Chern-Simons term to the gauge field action, and the resultant solution is horizonless,
regular and asymptotic to the extremal $BTZ$ black hole \cite{gc}.
Alternatively, one may introduce a dilaton field as a matter source coupled to
the Maxwell part $F^2$ in the action
(\ref{eq:EMaction}). For example,  by coupling the dilaton 
to the Maxwell part in the action by an exponential term, $e^{-4a\phi}F^2$ , static electrically
charged dilaton black hole solutions were obtained in \cite{cm}. The electric
potential in $g_{tt}$ and $g_{rr}$ is just a constant term and  as a result the quasilocal mass is finite at spatial infinity. One may also couple a dilaton to
the pure magnetic solution in \cite{hw} to attempt to get a finite mass solution. 
In recent, a family of spinning version of the uncharged dilaton black holes are obtained in
\cite{chaman}. Generalizations to charged cases are in progress.

\vspace{0.5cm}

The author would like to thank Eric  Hirschmann and Dean Welch for interesting discussions, and Robert Mann for his comments on the quasilocal formalism. 
This work was supported in part by the Natural Science and Engineering Research Council of Canada.

\end{document}